\documentclass[final,1p,times,twocolumn]{elsarticle}

\usepackage{amssymb}
\usepackage{amsmath}
\usepackage{xcolor}
\newcounter{bla}

\journal{Computer Physics Communications}

\begin{document}

\begin{frontmatter}



\title{MoireStudio: A Universal Twisted Electronic Structure Calculation Package}

\author[a]{Junxi Yu}
\author[a]{Yichen Liu}
\author[a]{Cheng-Cheng Liu\corref{author}}

\cortext[author] {Corresponding author.\\\textit{E-mail address:} ccliu@bit.edu.cn}
\address[a]{Centre for Quantum Physics, Key Laboratory of Advanced Optoelectronic Quantum Architecture and Measurement (MOE), School of Physics, Beijing Institute of Technology, Beijing 100081, China}

\begin{abstract}
Twistronics is an emerging and captivating field in condensed matter physics and material science. However, accurately and efficiently calculating the electronic structures of twisted systems remains a significant challenge. To address this, we have developed MoireStudio, a universal Python-based computational package for twisted electronic structures. Its functionalities include commensurate structure search, structure generation, parameterization, and construction for tight-binding models and continuum models, and the precise incorporation of full relaxation effects. The package is applicable to arbitrary combinations of two-dimensional materials, including rectangular lattices and heterostructures. User-friendly and easy to use, MoireStudio supports parallel large-scale computations, provides visualization capabilities, and offers interfaces with third-party software. It is poised to become a convenient and powerful tool for researchers in twistronics fields.
\\


\noindent \textbf{PROGRAM SUMMARY}

\begin{small}
\noindent
{\em Program Title:} MoireStudio \\
{\em Licensing provisions:} MIT license  \\
{\em Programming language:} Python  \\
{\em Nature of problem:} Twist and lattice mismatch form moiré superlattices with electronic structures highly sensitive to geometry, interlayer hybridization, and lattice relaxation. Performing a full calculation involves multiple interdependent steps—finding commensurate moiré lattice vectors, constructing moiré structures, incorporating relaxation effects, building large Hamiltonians, and computing band structures or topological quantities. These tasks are often handled by separate scripts with inconsistent conventions, complicating parameter scans, reproducibility, and portability across computing environments.  Moreover, performing these computations often involves a certain level of technical barrier.\\
{\em Solution method:} This package provides a unified JSON‑based command‑line driver that organizes calculations into structure, tight‑binding, and continuum $k\cdot p$ workflows, ensuring consistent conventions and recording metadata for reproducibility. Geometry modules identify moiré lattice vectors and type, and can include relaxation via Fourier‑based relaxation displacement fields. Tight‑binding workflows use monolayer tight-binding models and the Slater-Koster-like method to build moiré Hamiltonians, stored and processed efficiently in sparse formats. Continuum $k\cdot p$ workflows construct plane‑wave Hamiltonians with kinetic and coupling terms. Eigenproblems are solved via parallel dense diagonalization or iterative sparse solvers, enabling scalable band calculations. Results are output in standardized formats for visualization and further analysis.\\
\end{small}
\end{abstract}
\end{frontmatter}
\section{Introduction}
Two-dimensional (2D) layered materials forming moiré superlattices through interlayer twist angles or lattice mismatch have given risen to a series of novel quantum phenomena. In recent years, twistronics has become a vibrant research frontier in condensed matter physics and material science, spanning from strong correlations, superconductivity, and topology in magic-angle graphene to fractional Chern insulator in twisted transition metal dichalcogenide (TMD) bilayers~\cite{cao_correlated_2018,serlin_intrinsic_2020,cao_unconventional_2018,yankowitz_tuning_2019,lu_superconductors_2019,cai_signatures_2023,zengThermodynamicEvidenceFractional2023,xuObservationIntegerFractional2023,guo2024superconductivity,hao_electric_2021,kang_evidence_2024,lu_fractional_2024,regan_mott_2020,xie_tunable_2025}.

Currently, mainstream methods for studying moiré systems mainly include density functional theory (DFT), tight-binding (TB) model, and the $k\cdot p$ continuum model\cite{naik_ultraflatbands_2018,devakulMagicTwistedTransition2021,mao_transfer_2024a,suarezmorell_flat_2010,tramblydelaissardiere_localization_2010,lopesdossantos_graphene_2007a,bistritzer_moire_2011,wu_topological_2019,cao_initio_2021,li_tbplas_2023,ma_topological_2021,gao_tunable_2020}. However, due to the unique properties of moiré systems, their applications face various challenges. (1) DFT offers high accuracy and broad applicability, but its computational cost is high. Particularly, the novel properties of twisted systems often emerge in supercells containing thousands or even tens of thousands of atoms, making DFT calculations prohibitively expensive. (2) The TB model method provides an intuitive physical picture and a balance between accuracy and computational cost, and is well-established. However, it is more suitable for simple structures. Obtaining accurate TB parameters for real materials still requires DFT calculations and Wannier interpolation. (3) Since its proposal, the twisted continuum model method has been notable for its clever implementation and minimal computational cost. However, parameterization for the continuum model is often difficult, and currently, there are very few materials whose continuum models are widely accepted. Furthermore, using this model requires a certain understanding of twistronics, which somewhat hinders its integration with other fields. Another noteworthy method is combining deep learning with DFT to enable the calculation of twisted structures\cite{mao_transfer_2024a,liu_dpmoire_2025,liDeeplearningDensityFunctional2022,bao_deeplearning_2024,xie_lattice_2023}. This approach is promising but still under development, and its high demand for training data is also a challenge. These methods, each with its merits and limitations for specific applications, are currently not systematically integrated and suffer from a range of problems. Additionally, relaxation is a crucial factor affecting the electronic structure of twisted systems\cite{nam_lattice_2017,tarnopolsky_origin_2019,wang_fractional_2024,mao_transfer_2024a}, but the aforementioned methods struggle to provide an efficient and accurate solution for this problem. These issues more or less impede the development of the twistronics field. On the other hand, despite the vast number of 2D materials, only a tiny fraction of them have been studied within the framework of moir\'e superlattices, leaving numerous systems awaiting further exploration. 

Therefore, building upon our developed general method for calculating twisted electronic structures\cite{yu_general_2025} and the analytical relaxation framework for twisted systems\cite{yu_relaxation_2025}, we have developed an integrated software package for the calculation of twisted electronic structures. Its functionalities encompass commensurate structure search and generation, construction of twisted TB models and continuum models, and the precise and comprehensive consideration of relaxation effects. It is applicable to arbitrary 2D materials, significantly reduces computational cost, and strives to efficiently and accurately study the physical properties of twisted systems from different perspectives. It aims to provide a powerful and convenient research tool for the field.

This manuscript is organized as follows. In Section \ref{sec:method}, we provide a brief introduction to the theoretical methodology related to MoireStudio. In Section \ref{sec:feature}, we outline the operational logic and key features of MoireStudio, along with a user guide. Furthermore, in Section \ref{sec:examples}, we present several examples to demonstrate the versatility, efficiency, and user-friendliness of MoireStudio. Finally, in Section \ref{sec:discussion}, we provide a concise summary of the key features of MoireStudio, discuss its current limitations, and propose potential future directions.

\section{Methodology}\label{sec:method}
\subsection{Twisted Geometry}
When two stacked layers of 2D van der Waals materials have a relative rotation or lattice mismatch, the original translational symmetry can be broken, forming a larger lattice period known as a moiré pattern. It is important to emphasize that not all twist angle produces a moiré crystal; only at certain specific angles does the system possess complete translational symmetry, a condition we refer to as a commensurate condition.

We assume the top layer rotates by $-\frac{\theta}{2}$ and the bottom layer rotates by $\frac{\theta}{2}$. Denote the lattice vectors of the 2D material as $\pmb{a}_{i}^{l}$, where $i=\{1,2\}$, $l=\{t,b\}$ represent the top and bottom layers, respectively. We can define the commensurate condition for the twisted system as: there exists a set of integers $m_1,n_1,m_2,n_2$, not all zero, such that
\begin{equation}\label{eq:comsurate_condition}
    R(-\frac{\theta}{2})(m_{1}\pmb{a}_{1}^{t}+m_{2}\pmb{a}_{2}^{t})=
    R(\frac{\theta}{2})(n_{1}\pmb{a}_{1}^{b}+n_{2}\pmb{a}_{2}^{b}),
\end{equation}
holds, where $R(\pm \frac{\theta}{2})$ represents the rotation matrix for $\pm\frac{\theta}{2}$. It implies that the system is commensurate only when a common periodicity exists between the two rotated lattice layers. For clarity, we can write this in matrix form
\begin{equation}
    R(-\frac{\theta}{2})L_{t}
    \begin{pmatrix}
        m_{1}\\
        m_{2}
    \end{pmatrix}
    =
    R(\frac{\theta}{2})L_{b}
    \begin{pmatrix}
        n_{1}  \\
        n_{2}
    \end{pmatrix}.
\end{equation}
Here, $L_{l}=(\pmb{a}_{1}^{l},\pmb{a}_{2}^{l})$ is the 2D lattice matrix for layer $l$. Further transformation yields
\begin{equation}
    \begin{pmatrix}
        m_{1}  \\
        m_{2}
    \end{pmatrix}
    =
    L_{t}^{-1}R(\theta)L_{b}
    \begin{pmatrix}
        n_{1}  \\
        n_{2}
    \end{pmatrix}.
\end{equation}
The commensurate condition requires $m_1,n_1,m_2,n_2$ to be integers, not all zero. Therefore, the condition for the twisted structure to have complete translational symmetry is that the matrix $L_{t}^{-1}R(\theta)L_{b}$ is a rational matrix (irrational matrix correspond to quasicrystal cases, which will be considered in the next version of our package). We can transform this rationality problem into solving a Diophantine equation to identify whether a given twist angle $\theta$ is commensurate. Furthermore, once a commensurate angle is selected, the moir\'e lattice structure can be readily obtained through Eq.~\ref{eq:comsurate_condition}.

Based on the above theory, we can implement, for any 2D material combination with arbitrary lattice and constants (including rarely studied cases like rectangular lattices, heterostructures, etc.), the study of their commensurability and the generation of moiré atomic structures.

\subsection{Twisted TB Model}
The TB method is widely used in condensed matter physics due to its intuitive physical picture and efficient computational process. It also finds application in twistronics, but accurately constructing twisted TB models remains a challenge.

From the perspective of local stacking approximation, if we focus on a local part of the twisted structure, we can simply view it as an untwisted bilayer structure with a certain interlayer displacement, ignoring the local rotation. That is, the twisted structure can be approximated as a combination of all possible stacking configurations. In other words, the electronic structures of all possible untwisted bilayer configurations contain all the necessary information to construct their twisted counterparts. Additionally, due to symmetry considerations, these possible stacking configurations can be reduced to a finite set of distinct structures. Therefore, we only need to obtain the TB models for several stacking configurations of the bilayer structure via DFT-Wannier interpolation, and then construct the twisted TB model using the method below.

Within the TB framework, we divide the twisted TB Hamiltonian into the interlayer part and the intralayer part.
Interlayer coupling is the core and also the key difficulty of constructing the twisted Hamiltonian, as it often dominates the most interesting properties in twisted systems. Fortunately, interlayer coupling is a van der Waals interaction, which is weak enough that we can adopt the two-center approximation\cite{jung_initio_2014}. This assumes that the interlayer coupling strength between two orbitals $i, j$ depends only on the distance between them. Thus, the interlayer coupling function can be written as\cite{yu_general_2025}
\begin{equation}
    t_{ij}(\pmb{r})=h_{0}^{ij}\exp{[-\frac{l(r_{z}-d_{0}^{ij})^{2}}{(r_{0}^{ij})^{2}}]}
    \exp{[-\frac{r_{x}^{2}+r_{y}^{2}}{(r_{0}^{ij})^{2}}]}.
\end{equation}
Here, $\pmb{r}=(r_{x},r_{y},r_{z})$ is the distance between the $i$-th orbital in the top layer and the $j$-th orbital in the bottom layer, $l$ is the sign of $r_{z}-d_{0}^{ij}$. $h_{0}^{ij}, d_{0}^{ij}, r_{0}^{ij}$ are parameters to be determined, representing the coupling strength, interlayer distance, and decay distance between these two orbitals, respectively. They can be accurately solved by substituting the Hamiltonians of several non-equivalent stacking configurations. Thus, after obtaining the precise coupling function and combining it with the twisted geometry, the interlayer part of the twisted Hamiltonian can be obtained.

Similarly, the intralayer Hamiltonian can also be obtained from DFT-Wannier interpolation calculations on the monolayer structure. Generally, the difference in the intralayer Hamiltonian before and after twisting can be considered negligible, so the twisted intralayer Hamiltonian can be directly taken as the untwisted case. Additionally, when the original bilayer system exhibits electrical polarization, the diagonal part of the intralayer Hamiltonian might be slightly affected by the stacking configuration\cite{wu_topological_2019,zhang_polarizationdriven_2024}. Nevertheless, we can always extract sufficient intralayer information from the Hamiltonians of several stacking bilayer structures.

Through the above method, we can accurately obtain the twisted TB Hamiltonian. Compared to other methods, this approach focuses not only on the low-energy part but also on the overall electronic properties, with a manageable computational cost, making it more suitable for calculating large-twist-angle systems.

\subsection{Twisted Continuum Model}
The most widely used model in theoretical studies of twistronics is the continuum model. This method, through a clever plane-wave expansion combined with low-energy $k\cdot p$ theory, can accurately reproduce the low-energy electronic structure of twisted systems with very few parameters and minimal computational cost. MoireStudio also incorporates this part.

Starting from low-energy $k\cdot p$ theory, we can obtain the monolayer Hamiltonian $h^{l}(\pmb{k})$. We can use effective mass, Fermi velocity, or other forms of a $k\cdot p$ effective model to describe the low-energy electrons of the monolayer system, typically applicable near a high-symmetry point in the Brillouin zone. For the interlayer coupling and intralayer potential, we perform a Fourier transform to describe them. For the untwisted bilayer structure under in-plane interlayer displacement $\pmb{\delta}$, the Hamiltonian becomes
\begin{equation}
    H(\pmb{\delta},\pmb{k})=
    \begin{pmatrix}
        h^{t}(\pmb{k})+V^{t}(\pmb{\delta}) & T(\pmb{\delta}) \\
        T^{\dagger}(\pmb{\delta}) & h^{b}(\pmb{k})+V^{b}(\pmb{\delta})
    \end{pmatrix}.
\end{equation}
$V^{l}(\pmb{\delta})$ is the response of the intralayer Hamiltonian to the interlayer displacement, and $T(\pmb{\delta})$ is the interlayer coupling. Both can be obtained by fitting the Hamiltonians of several bilayer structures with different stacking configurations (corresponding to interlayer displacements $\pmb{\delta}$).

From the local stacking approximation perspective, the local displacement can be considered entirely due to the twist: $\pmb{\delta}=[R(\frac{\theta}{2})-R(-\frac{\theta}{2})]\pmb{r}$. Therefore, $\pmb{G}_{j}\cdot\pmb{\delta}=\pmb{g}_{j}\cdot\pmb{r}$, and we obtain the continuum model Hamiltonian\cite{lopesdossantos_graphene_2007a,bistritzer_moire_2011,wu_topological_2019}
\begin{equation}\label{eq:continuum_model}
    H(\pmb{r})=
    \begin{pmatrix}
        h^{t}(-i\pmb{\nabla})+V^{t}(\pmb{r}) &  T(\pmb{r}) \\
        T^{\dagger}(\pmb{r}) & h^{b}(-i\pmb{\nabla})+V^{b}(\pmb{r})
    \end{pmatrix}.
\end{equation}
This is the most commonly used twisted continuum model, suitable for small-angle twisted systems. Here, $\pmb{G}_{j}$ are the reciprocal lattice vectors of the untwisted bilayer structure. We define $\pmb{G}_{0}=0, \pmb{G}_{1}=\frac{4\pi}{\sqrt{3}a_{0}}$, and number all reciprocal lattice points in order of increasing distance counterclockwise. $\pmb{g}_{j}$ are the corresponding moiré reciprocal lattice vectors.

\subsection{Relaxation}
Relaxation also plays a crucial role in twisted systems, profoundly affecting both the band morphology and topological properties. Incorporating our recently developed method\cite{yu_relaxation_2025}, we can accurately and completely address the relaxation problem in twisted systems.

In twisted systems, relaxation can be decomposed into in-plane and out-of-plane components. Starting from elastic theory and solving the Euler-Lagrange equations, we can obtain the relaxation displacement field for the twisted system. Taking a hexagonal lattice as an example, the in-plane and out-of-plane relaxations are
\begin{equation}\label{eq:relax_field}
    \begin{aligned}
        \pmb{u}_{\parallel}^{-}(\pmb{r}) &= \frac{1}{\theta^{2}}\kappa_{\parallel}\sum_{j\in \{1,3,5\}}\frac{1}{|\pmb{G}_{j}|^{2}}\sin(\pmb{g}_{j}\cdot \pmb{r})\pmb{G}_{j},\\
        \pmb{u}_{\perp}^{-}(\pmb{r}) &=\frac{1}{\theta^{2}}\kappa_{\perp}\sum_{j\in \{1,3,5\}}\frac{1}{|\pmb{G}_{j}|^{2}}\cos(\pmb{g}_{j}\cdot\pmb{r})\hat{\pmb{e}}_{z},
    \end{aligned}
\end{equation}
here, $\pmb{u}_{\parallel/\perp}^{-}(\pmb{r})=\pmb{u}_{\parallel/\perp}^{t}(\pmb{r})-\pmb{u}_{\parallel/\perp}^{b}(\pmb{r})$ is the difference between the relaxation displacement fields of the top and bottom layers, and $\kappa_{\parallel},\kappa_{\perp}$ are coefficients characterizing the relaxation strength. The form of out-of-plane relaxation above can also be obtained by performing z-direction relaxation on bilayer structures with different interlayer displacements using the DFT method.

The question now is how relaxation affects the electronic structure. For the TB model, once the exact functional forms of interlayer coupling and on-site energy dependence on interlayer displacement are obtained, simply substituting the relaxed atomic structure yields the precise twisted Hamiltonian. For the continuum model, the problem is more complex. Briefly, out-of-plane relaxation introduces a correction to the strength coefficients of interlayer coupling and intralayer moiré potential, while in-plane relaxation acts like a gauge field, introducing an additional phase factor to the coupling terms. We can expand this phase factor, mapping the in-plane relaxation to higher-order coupling effects in reciprocal space, thereby fully accounting for the impact of relaxation on the electronic structure\cite{yu_relaxation_2025}.

\section{Package features}\label{sec:feature}
\subsection{Overview}
\begin{figure}
    \centering
    \includegraphics[width=1\linewidth]{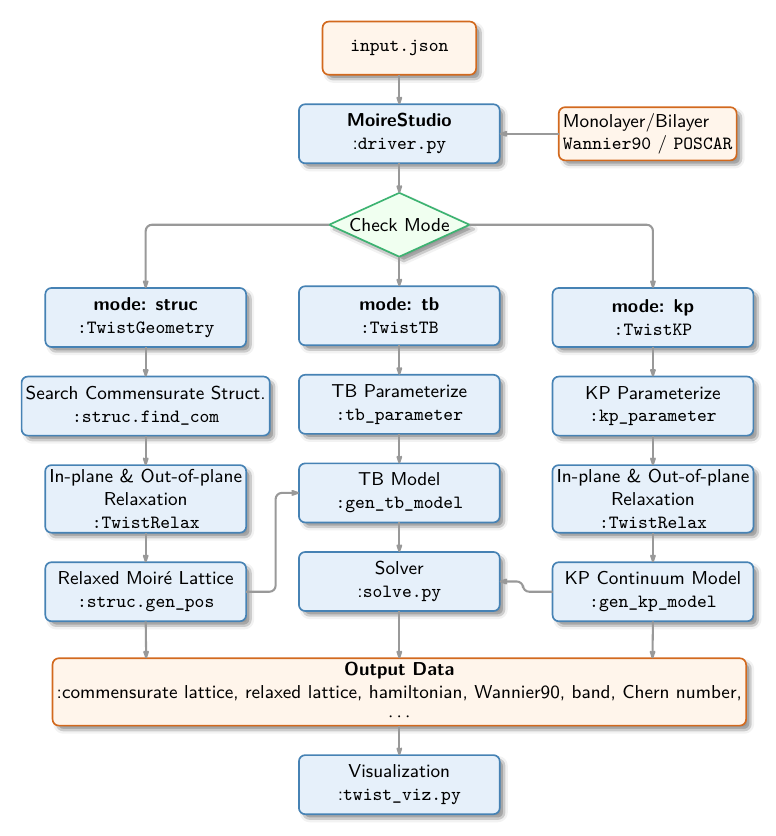}
    \caption{Workflow of MoireStudio. Starting from an input.json file, the system constructs and analyzes twisted material systems from three perspectives: structural geometry, TB model, and continuum model. It outputs results such as Hamiltonians, band structures, and topological properties. MoireStudio also supports visualization and interfacing with other computational software.}
    \label{fig:workflow}
\end{figure}

Starting from the user-provided instruction file \verb|input.json| together with external data files (e.g., Wannier data and POSCAR), MoireStudio can automatically execute three distinct computational workflows as shown in Fig.~\ref{fig:workflow}. In the structure-processing workflow, MoireStudio first searches for commensurate twist angles of a given bilayer lattice (or heterostructure). After selecting a target commensurate angle, it generates the corresponding twisted atomic structure, optionally incorporating both in-plane and out-of-plane lattice relaxation. 

In the TB workflow, MoireStudio takes a set of untwisted bilayer Hamiltonians and parameterizes them into intralayer Hamiltonian and interlayer coupling functions for twisted configurations. Combined with the relaxed atomic coordinates, this procedure yields an accurate TB model for the twisted system. 

In the continuum $k\cdot p$ workflow, MoireStudio constructs the continuum model using given parameters, and then incorporates our relaxation theory for twisted systems to construct an accurate relaxation-enabled continuum model. 

Based on the resulting models, MoireStudio supports the selection of appropriate solvers (such as serial, parallel, and/or sparse-matrix approaches) to compute energy spectra and related quantities. It outputs band-structure data, topological properties, relaxed lattice structures, and Wannier90 format data for straightforward interfacing with other software packages. Finally, MoireStudio also provides basic visualization utilities.

Below, we provide a detailed introduction to the specific implementation methods of each module.

\subsection{Installation and Operation}
MoireStudio requires \texttt{Python} \(\geq 3.10\) and depends only on three widely used scientific packages:
\texttt{numpy} \(\geq 1.23\), \texttt{scipy} \(\geq 1.9\), and \texttt{matplotlib} \(\geq 3.7\). MoireStudio can be installed via \texttt{pip} as
\begin{verbatim}
  pip install MoireStudio
\end{verbatim}
For offline installation, users may alternatively install from the wheel file
\begin{verbatim}
  pip install moirestudio-1.0.0-py3-none-any.whl
\end{verbatim}

After installation, MoireStudio can be executed from any directory containing an \texttt{input.json} file by running
\begin{verbatim}
  MoireStudio
\end{verbatim}
in the command line.

\subsection{Inputs and outputs}
In MoireStudio, there are three types of input files.
Primarily: \verb|input.json|, wannier.data, POSCAR.

\verb|input.json| is the essential control file for the package, governing calculation mode, parameters, parallelism, etc. Specific control parameters will be introduced in the Features and Examples sections, and can also be found in README.md.

Wannier data is necessary for parameterization and TB calculations. Specifically, these are wannier90.win, wannier90\_hr.dat, and wannier90\_centre.xyz, which are output files from the software Wannier90\cite{marzariMaximallyLocalizedGeneralized1997,mostofiWannier90ToolObtaining2008,souzaMaximallyLocalizedWannier2001}. The latter two can be obtained by selecting the following options in a Wannier90 calculation
\begin{verbatim}
    write_hr = True
    write_xyz = True
\end{verbatim}
For reading and writing these files, we have adopted the data processing approach used in the PythTB package\cite{Cole_Python_Tight_Binding_2025}.

POSCAR is a structure information file in VASP format and is an optional input\cite{kresse_efficient_1996}. When the calculation task only involves the geometry structure module, wannier90 data is not needed; only POSCAR is required, specified in \verb|input.json|.

The output files include TB model (wannier90\_hr.dat), band data (BAND.dat in vaspkit format\cite{wang_vaspkit_2021}), Chern number (Chern\_num.txt), Hamiltonian (twist\_ham.npy), commensurate angle list (commensurate\_angle.txt), lattice structure (POSCAR), as well as band structure plots in various formats.

\subsection{Geometry Module}
The geometry module has two main functions, controlled by 

\begin{verbatim}
  "task": "find_com" | "gen_pos"
\end{verbatim}
finding commensurate angles and generating structures.
Here is an example for searching commensurate angles in a homostructure:
\begin{verbatim}
{
  "mode": "struc",
  "task": "find_com",
  "struc": {
    "read_mode": "POSCAR",
    "read_layer": "bilayer",
    "ihetero": false,
    "d_0": 3.5,
    "max_angle": 180.0,
    "min_angle": 1,
    "lim": 20,
    "step": 0.02,
    "accurate": 0.01
  }
}
\end{verbatim}
Here, read\_mode indicates the file format, which can be POSCAR or Wannier90 format. ihetero=false indicates a homostructure. read\_layer=bilayer means the input file is for a bilayer structure. max/min\_angle defines the search range for commensurate angles. The step determines the search step size. accurate indicates the search precision; generally, higher precision requires more computation but is typically manageable for single-core serial execution. lim is the real-space search range, requiring lim multiplied by the lattice constant to be larger than the expected moiré unit cell lattice constant. For heterostructures, set ihetero to true and change read\_layer to monolayer. In this case, provide the top and bottom layer POSCAR files named t\_POSCAR and b\_POSCAR, respectively. This function outputs a file commensurate\_angle.txt. The first column is the angle, the second column is a metric for the commensurability (larger number indicates smaller moiré unit cell), and the third column is the estimated approximate number of atoms in the moiré unit cell. 

On the other hand, an example for generating a moiré structure is
\begin{verbatim}
{
  "mode": "struc",
  "task": "gen_pos",
  "theta_deg": 21.79,
  "struc": {
    "read_mode": "POSCAR",
    "read_layer": "bilayer",
    "ihetero": false,
    "d_0": 3.5,
    "zero_point": [0, 0],
    "lim": 20,
    "accurate": 0.01,
    "if_gen_pos": true
  }
}
\end{verbatim}
The twist angle is input via the theta\_deg entry. zero\_point indicates the rotation center; [0,0] means rotating around the first atom of the top layer and the first atom of the bottom layer. If input as a 2D array, e.g., [[0.0, 0.0],[0.5, 0.5]], it means rotating around the fractional coordinates (0, 0) of the top layer and (0.5, 0.5) of the bottom layer.

\subsection{Tight-Binding Module}
The TB module in our package is powerful, capable of automated parameter calculation, supports magnetic and spin-orbit coupled systems, allows adding electric fields, and considers relaxation. Based on system size, we designed three computation modes: fast serial, normal parallel, and sparse matrix parallel. Here is an example \verb|input.json|
\begin{verbatim}
{
  "mode": "tb",
  "task": "band",
  "theta_deg": 21.79,
  "struc": {
    "read_mode": "wannier",
    "read_layer": "bilayer",
    "ihetero": false,
    "d_0": 3.5
  },
  "tb": {
    "cores": 4,
    "isparse": false,
    "k_path": [[0,0],[0.5,0],[0.3333333,0.3333333],[0,0]],
    "nk": 201
  }
}
\end{verbatim}
This will output the Hamiltonian, band structure, and k-path information. cores controls the number of parallel cores. isparse controls whether to enable sparse matrix mode. 

\subsection{Continuum Model Module}
In the continuum model module, due to the fewer parameters, we support both direct parameter input and parameterization from wannier data. It supports outputting band structures, Hamiltonians, and Chern numbers.
Here is an example for twisted MoTe$_2$:
\begin{verbatim}
{
  "mode": "kp",
  "task": "band",
  "theta_deg": 1.2,
  "kp": {
    "tr": 4,
    "valley_pos": [0.6666667, 0.3333333],
    "mono_lat": [[3.46, 0.0], [1.73, 2.996447897094158]],
    "mass": [0.62,0.62],
    "valley_name": "K",
    "V_z": 0.0,
    "m_z": 0.0,
    "inter_idxs": [[0,0],[-1,-1],[-1,0]],
    "inter_amps": [-8.5, -8.5, -8.5],
    "inter_phas": [0, 0, 0],
    "intra_t_idxs": [[1,0],[1,1],[0,1],[-1,0],[-1,-1],[0,-1]],
    "intra_t_amps": [8, 8, 8, 8, 8, 8],
    "intra_t_phas": [-89.6, 89.6, -89.6, 89.6, -89.6, 89.6],
    "intra_b_idxs": [[1,0],[1,1],[0,1],[-1,0],[-1,-1],[0,-1]],
    "intra_b_amps": [8, 8, 8, 8, 8, 8],
    "intra_b_phas": [89.6, -89.6, 89.6, -89.6, 89.6, -89.6],
    "k_path": [[0,0],[0.666666,0.333333],[0.5,0.5],[0,0]],
    "nk": 201,
    "labels": ["G","K","M","G"],
    "ylim": [0, 25]
  }
}
\end{verbatim}
In the above content, idxs indicate coupling vectors, expressed in terms of moiré reciprocal lattice vectors, and tr is the truncation for the continuum model. It should be emphasized that in the above content
\begin{verbatim}
  "inter_idxs": [[i_11,i_12],[i_21,i_22],[i_31,i_32]],
  "inter_amps": [a_1, a_2, a_3],
  "inter_phas": [p_1, p_2, p_3]
\end{verbatim}
corresponding to 
\begin{equation}
  T(\pmb{r})=a_{1}e^{ip_{1}}e^{i(i_{11}\pmb{g}_{1}+i_{12}\pmb{g}_{2})\cdot \pmb{r}}+a_{2}e^{ip_{2}}e^{i(i_{21}\pmb{g}_{1}+i_{22}\pmb{g}_{2})\cdot \pmb{r}}+a_{3}e^{ip_{3}}e^{i(i_{31}\pmb{g}_{1}+i_{32}\pmb{g}_{2})\cdot \pmb{r}},
\end{equation}
in Eq.~(\ref{eq:continuum_model}), here, $(\pmb{g}_{1}, \pmb{g}_{2})$ is moir\'e reciprocal lattice vectors. The same applies to intralayer potential.

\section{Examples}\label{sec:examples}
\subsection{Geometry Module}
\begin{figure}
    \centering
    \includegraphics[width=1\linewidth]{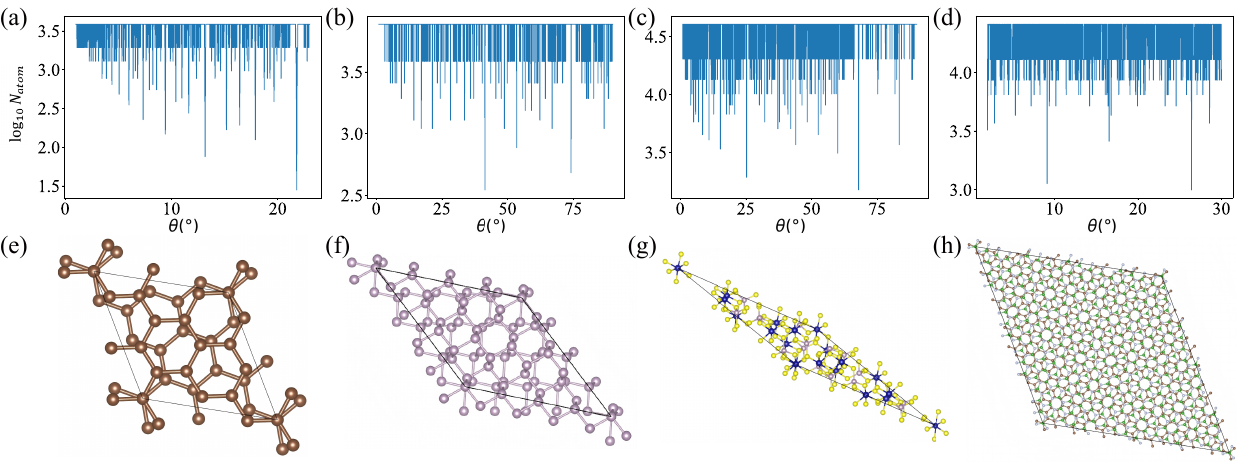}
    \caption{Example of MoireStudio structural module.(a)-(d) twisted bilayer graphene, twisted black phosphorene, twisted CrPS$_{4}$, and graphene-hexagonal boron nitride (hBN) heterostructures. Here, graphene and hBN have similar lattice vectors essentially, but their lattice constants differ by about 2\%. The abscissa represents the angle, while the ordinate represents log$_{10} N_{atom}$, where $N_{atom}$ is the number of atoms in the primitive cell corresponding to the given commensurate angle. (e)-(h) The minimum commensurable structure of each system.}
    \label{fig:struc}
\end{figure}

In this section, we present four examples to demonstrate the commensurate-angle search and structure generation capabilities of MoireStudio. These examples are twisted bilayer graphene, twisted black phosphorene, twisted CrPS$_{4}$, and graphene-hexagonal boron nitride (hBN) heterostructures. They represent hexagonal lattices, rectangular lattices, complex low-symmetry structures, and heterostructure configurations, respectively, as illustrated in Fig.~\ref{fig:struc}.
In our package, setting \verb|input.json| as
\begin{verbatim}
  {
 "mode":"struc",
	"task": "find_com",      
	"struc":{
		"read_mode": "POSCAR",
		"step": 0.01,
		"lim":30,
		"accurate":5e-3,
		"max_angle": 23,
		"min_angle": 1
  }
  }
\end{verbatim}
yields the distribution of all commensurate angles, as shown in Figs.~\ref{fig:struc}(a)-(d). Selecting one of these angles and setting \verb|task='gen_pos'| gives the corresponding commensurate structure, as shown in Figs.~\ref{fig:struc}(a)-(d). 
Set \verb|"ihetero": true| to perform heterostructures calculations.

\begin{figure}
    \centering
    \includegraphics[width=1\linewidth]{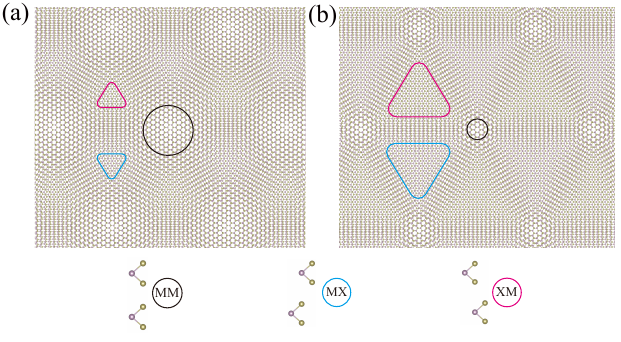}
    \caption{The influence of relaxation on the structure of tTMD. (a) Unrelaxed structure; (b) relaxed structure.}
    \label{fig:tmd_relaxed_struc}
\end{figure}

Furthermore, MoireStudio also supports the generation of fully relaxed structures by adding the content in the \verb|input.json| file as
\begin{verbatim}
"relax": {
"irelax": true,
"kappa_parallel": 2e-3,
"order_num": 1,
"kappa_perp": 1e-3
}
\end{verbatim}
Taking tTMDs as an example, a comparison of their lattice structures before and after relaxation is shown in Fig.~\ref{fig:tmd_relaxed_struc}. It can be observed that the relaxation results obtained from our elastic continuum mechanics approach are consistent with those previously obtained via DFT and molecular dynamics simulation\cite{mao_transfer_2024a,zhangUniversalMoireModelBuildingMethod2024a,jia_moire_2024a}.

\subsection{TB Module}
Next, we apply the TB module of MoireStudio to several typical systems. First, for the extensively studied twisted bilayer graphene (TBG) and twisted MoS$_{2}$, at a twist angle of $\theta=21.79^{\circ}$, we use the bilayer Wannier90 output files for both AA- and AB-stacked configurations as input data. The \verb|input.json| is configured as
\begin{verbatim}
  {
  "mode": "tb",
  "task": "band",
  "theta_deg": 21.79,
  "struc": {
    "read_mode": "WANNIER",
    "read_layer": "bilayer",
    "bilayer_num": 2,
    "ihetero": false,
    "d_0": 3.60
  },
  "tb": {
    "cores": 1,
    "isparse": false,
    "k_path": [[0,0,0],[0.333333,0.333333,0],[0.5,0.0,0],[0,0,0]],
    "nk": 121
  }
}
\end{verbatim}
\begin{figure}
    \centering
    \includegraphics[width=1\linewidth]{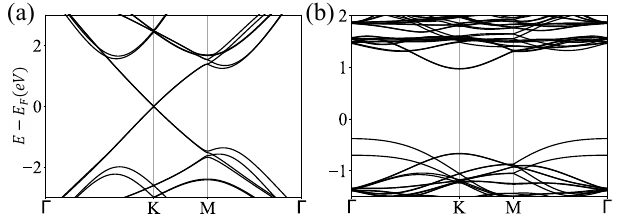}
    \caption{Band structure of (a) TBG and (b) twisted MoS$_{2}$ with twist angle $\theta=21.79^{\circ}$ calculated by TB module of MoireStudio.}
    \label{fig:tb_tbg_mos2}
\end{figure}
The resulting band structure is shown in Fig.~\ref{fig:tb_tbg_mos2}. 

\begin{figure}
    \centering
    \includegraphics[width=1\linewidth]{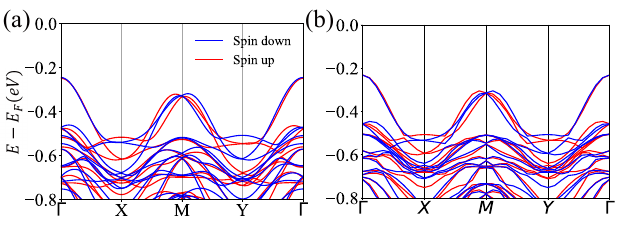}
    \caption{Band structure of twisted bilayer CrPS$_{4}$ at a twist angle of $\theta=67.92^{\circ}$: (a) calculated by MoireStudio, (b) calculated by DFT (VASP)\cite{kresse_efficient_1996}.}
    \label{fig:tb_crps4}
\end{figure}

Furthermore, MoireStudio also supports calculations for magnetic complex systems. Taking CrPS$_4$ as an example, we select a twist angle of $\theta=67.92^{\circ}$ and obtain the twisted bilayer band structure. A comparison with DFT results is presented in Fig.~\ref{fig:tb_crps4}, demonstrating a high degree of consistency between the two methods. It can be seen that due to broken spatial inversion and in-plane $C_{2x}$ symmetry connecting different spin sublattices, twisted CrPS$_4$ becomes an altermagnet\cite{liu_twisted_2024a}.

\begin{figure}
    \centering
    \includegraphics[width=1\linewidth]{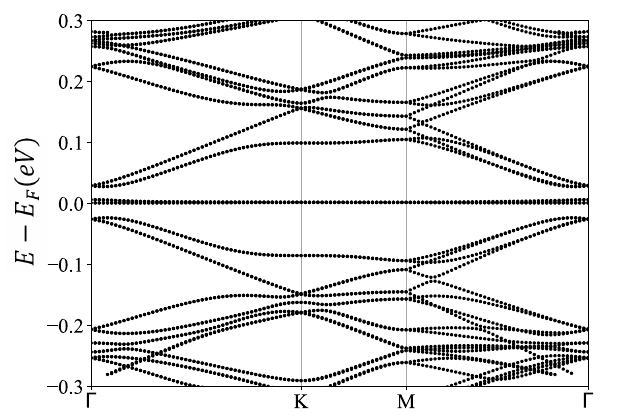}
    \caption{Band structure of magic-angle($1.05^{\circ}$) TBG calculated using the parallel sparse matrix scheme in the TB module of MoireStudio with full relaxation.}
    \label{fig:tb_tbg}
\end{figure}

In addition, using the TB method to compute small-angle twisted systems with particularly large structural size also demands substantial computational resources. Therefore, we also provide a parallelization scheme. Users can enable parallel computing by setting the \verb|cores| to the desired number of parallel cores. For Hamiltonian generation and diagonalization, the default uses dense diagonalization. For extremely large systems, it is recommended to enable the \verb|isparse| to employ a parallel iterative sparse matrix solver. Taking magic-angle (1.05$^{\circ}$) TBG as an example, whose unit cell contains over ten thousand atoms, we used 64 parallel cores combined with the sparse matrix solver. This approach required only approximately ten minutes to obtain the band structure, as shown in Fig.~\ref{fig:tb_tbg}, which is also consistent with experimental and DFT results\cite{koshino_maximally_2018,jung_initio_2014}.

\subsection{Continuum Model Module}

\begin{figure}
\centering
\includegraphics[width=1\linewidth]{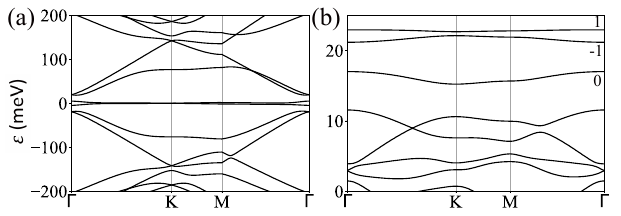}
\caption{Band structures calculated using the continuum model module of MoireStudio. (a) Band structure of TBG at $\theta=1.05^{\circ}$ with relaxation; (b) band structure of tMoTe$_{2}$ at $\theta=1.2^{\circ}$. The numbers near the bands are their corresponding Chern numbers.}
\label{fig:kp_tbg_tmd}
\end{figure}

In the continuum model section, we demonstrate the capabilities of MoireStudio using two highly studied twisted systems, TBG and tMoTe$_{2}$, as examples. By setting \verb|"task": "band"| in \verb|input.json|, band structure calculations can be performed; similarly, setting \verb|"task": "chern"| enables the calculation of Chern numbers. The results, shown in Fig.~\ref{fig:kp_tbg_tmd}, are consistent with previous findings\cite{koshino_maximally_2018,wu_topological_2019}.

\begin{figure}
\centering
\includegraphics[width=1\linewidth]{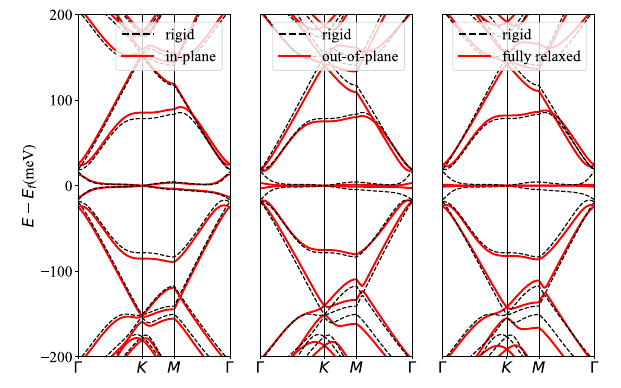}
\caption{Effects of progressively incorporating relaxation on the low-energy bands of magic-angle TBG: (a) in-plane, (b) out-of-plane, and (c) combined relaxations.}
\label{fig:kp_tbg_relax}
\end{figure}

In the kp continuum model, we can also consider relaxation. According to the theory in Ref.~\cite{yu_relaxation_2025}, we can treat out-of-plane relaxation as a correction to the coupling coefficients, while in-plane relaxation provides an additional phase factor, leading to longer-range couplings. We studied the effects of under different relaxation conditions on the electronic structure of magic-angle TBG as shown in Fig.~\ref{fig:kp_tbg_relax}. Since both in-plane and out-of-plane relaxation reduce the contribution from AA stacking regions and enhance the contribution from AB stacking regions, relaxation in TBG more readily promotes the formation of isolated flat bands. These results are also consistent with previous findings\cite{cantele_structural_2020,lisi_observation_2021,lucignano_crucial_2019,kazmierczak_strain_2021}。

\begin{figure}
\centering
\includegraphics[width=1\linewidth]{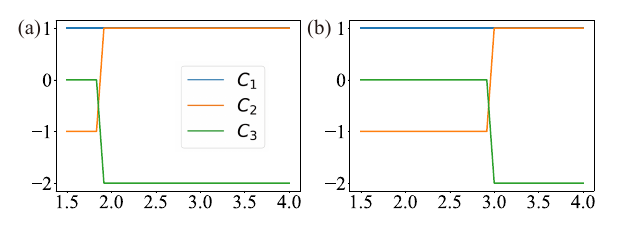}
\caption{Change in the topology of the three highest valence bands of tMoTe$_{2}$ with twist angle, shown (a) without and (b) with lattice relaxation.}
\label{fig:kp_tmd_relax}
\end{figure}

Furthermore, relaxation profoundly affects the topological properties of twisted systems. The phase diagram of the Chern numbers for the first three bands in tMoTe$_2$ as a function of twist angle is shown in Fig.~\ref{fig:kp_tmd_relax}. The results considering relaxation are consistent with DFT calculations and experiments, demonstrating that relaxation deeply influences the electronic properties of the system\cite{jia_moire_2024a,wang_fractional_2024,mao_transfer_2024a}.

\section{Discussion}\label{sec:discussion}
In summary, we have presented a comprehensive universal computational package for moiré twisted systems. Its functionalities cover geometry, TB, and continuum model, fully incorporate relaxation effects, support large-scale parallel computation, and provide interfaces with other software and visualization tools. We have also provided a user guide and numerous examples representing different scenarios. We believe MoireStudio will become a powerful tool in related fields.

In subsequent releases, we also plan to extend MoireStudio toward many-body and transport capabilities, including electron–electron interaction calculations (e.g., Hartree–Fock mean-field theory and exact diagonalization) and transport-related properties such as magneto-optical responses and various Hall effects.

\section*{Acknowledgments}
The work is supported by the NSF of China (Grant No. 12374055), the Science Fund for Creative Research Groups of NSFC (Grant No. 12321004), and the National Key R\&D Program of China (Grant No. 2020YFA0308800).

\bibliographystyle{elsarticle-num}
\bibliography{ref.bib}

@article{cao_correlated_2018,
  title = {Correlated Insulator Behaviour at Half-Filling in Magic-Angle Graphene Superlattices},
  author = {Cao, Yuan and Fatemi, Valla and Demir, Ahmet and Fang, Shiang and Tomarken, Spencer L. and Luo, Jason Y. and {Sanchez-Yamagishi}, Javier D. and Watanabe, Kenji and Taniguchi, Takashi and Kaxiras, Efthimios and Ashoori, Ray C. and {Jarillo-Herrero}, Pablo},
  year = {2018},
  journal = {Nature},
  volume = {556},
  number = {7699},
  pages = {80--84},
  doi = {10.1038/nature26154}
}

@article{serlin_intrinsic_2020,
  title = {Intrinsic Quantized Anomalous {{Hall}} Effect in a Moir{\'e} Heterostructure},
  author = {Serlin, M. and Tschirhart, C. L. and Polshyn, H. and Zhang, Y. and Zhu, J. and Watanabe, K. and Taniguchi, T. and Balents, L. and Young, A. F.},
  year = {2020},
  journal = {Science},
  volume = {367},
  number = {6480},
  pages = {900--903},
  doi = {10.1126/science.aay5533}
}

@article{cao_unconventional_2018,
  title = {Unconventional Superconductivity in Magic-Angle Graphene Superlattices},
  author = {Cao, Yuan and Fatemi, Valla and Fang, Shiang and Watanabe, Kenji and Taniguchi, Takashi and Kaxiras, Efthimios and {Jarillo-Herrero}, Pablo},
  year = {2018},
  journal = {Nature},
  volume = {556},
  number = {7699},
  pages = {43--50},
  doi = {10.1038/nature26160}
}

@article{yankowitz_tuning_2019,
  title = {Tuning Superconductivity in Twisted Bilayer Graphene},
  author = {Yankowitz, Matthew and Chen, Shaowen and Polshyn, Hryhoriy and Zhang, Yuxuan and Watanabe, K. and Taniguchi, T. and Graf, David and Young, Andrea F. and Dean, Cory R.},
  year = {2019},
  journal = {Science},
  volume = {363},
  number = {6431},
  pages = {1059--1064},
  doi = {10.1126/science.aav1910}
}

@article{lu_superconductors_2019,
  title = {Superconductors, Orbital Magnets and Correlated States in Magic-Angle Bilayer Graphene},
  author = {Lu, Xiaobo and Stepanov, Petr and Yang, Wei and Xie, Ming and Aamir, Mohammed Ali and Das, Ipsita and Urgell, Carles and Watanabe, Kenji and Taniguchi, Takashi and Zhang, Guangyu and Bachtold, Adrian and MacDonald, Allan H. and Efetov, Dmitri K.},
  year = {2019},
  journal = {Nature},
  volume = {574},
  number = {7780},
  pages = {653--657},
  doi = {10.1038/s41586-019-1695-0}
}

@article{cai_signatures_2023,
  title = {Signatures of Fractional Quantum Anomalous {{Hall}} States in Twisted {{MoTe2}}},
  author = {Cai, Jiaqi and Anderson, Eric and Wang, Chong and Zhang, Xiaowei and Liu, Xiaoyu and Holtzmann, William and Zhang, Yinong and Fan, Fengren and Taniguchi, Takashi and Watanabe, Kenji and Ran, Ying and Cao, Ting and Fu, Liang and Xiao, Di and Yao, Wang and Xu, Xiaodong},
  year = {2023},
  journal = {Nature},
  volume = {622},
  number = {7981},
  pages = {63--68},
  doi = {10.1038/s41586-023-06289-w}
}

@article{zengThermodynamicEvidenceFractional2023,
  title = {Thermodynamic Evidence of Fractional {{Chern}} Insulator in Moir{\'e} {{MoTe2}}},
  author = {Zeng, Yihang and Xia, Zhengchao and Kang, Kaifei and Zhu, Jiacheng and Kn{\"u}ppel, Patrick and Vaswani, Chirag and Watanabe, Kenji and Taniguchi, Takashi and Mak, Kin Fai and Shan, Jie},
  year = {2023},
  journal = {Nature},
  volume = {622},
  number = {7981},
  pages = {69--73},
  doi = {10.1038/s41586-023-06452-3}
}

@article{xuObservationIntegerFractional2023,
  title = {Observation of {{Integer}} and {{Fractional Quantum Anomalous Hall Effects}} in {{Twisted Bilayer MoTe}} 2},
  author = {Xu, Fan and Sun, Zheng and Jia, Tongtong and Liu, Chang and Xu, Cheng and Li, Chushan and Gu, Yu and Watanabe, Kenji and Taniguchi, Takashi and Tong, Bingbing and Jia, Jinfeng and Shi, Zhiwen and Jiang, Shengwei and Zhang, Yang and Liu, Xiaoxue and Li, Tingxin},
  year = {2023},
  journal = {Phys. Rev. X},
  volume = {13},
  number = {3},
  pages = {031037},
  doi = {10.1103/PhysRevX.13.031037}
}

@article{guo2024superconductivity,
  title = {Superconductivity in 5.0{$^\circ$} Twisted Bilayer {{WSe2}}},
  author = {Guo, Yinjie and Pack, Jordan and Swann, Joshua and Holtzman, Luke and Cothrine, Matthew and Watanabe, Kenji and Taniguchi, Takashi and Mandrus, David G. and Barmak, Katayun and Hone, James and Millis, Andrew J. and Pasupathy, Abhay and Dean, Cory R.},
  year = {2025},
  journal = {Nature},
  volume = {637},
  number = {8047},
  pages = {839--845},
  doi = {10.1038/s41586-024-08381-1}
}

@article{hao_electric_2021,
  title = {Electric Field--Tunable Superconductivity in Alternating-Twist Magic-Angle Trilayer Graphene},
  author = {Hao, Zeyu and Zimmerman, A. M. and Ledwith, Patrick and Khalaf, Eslam and Najafabadi, Danial Haie and Watanabe, Kenji and Taniguchi, Takashi and Vishwanath, Ashvin and Kim, Philip},
  year = {2021},
  journal = {Science},
  volume = {371},
  number = {6534},
  pages = {1133--1138},
  publisher = {American Association for the Advancement of Science},
  doi = {10.1126/science.abg0399}
}

@article{kang_evidence_2024,
  title = {Evidence of the Fractional Quantum Spin {{Hall}} Effect in Moir{\'e} {{MoTe2}}},
  author = {Kang, Kaifei and Shen, Bowen and Qiu, Yichen and Zeng, Yihang and Xia, Zhengchao and Watanabe, Kenji and Taniguchi, Takashi and Shan, Jie and Mak, Kin Fai},
  year = {2024},
  journal = {Nature},
  volume = {628},
  number = {8008},
  pages = {522--526},
  publisher = {Nature Publishing Group},
  doi = {10.1038/s41586-024-07214-5}
}

@article{lu_fractional_2024,
  title = {Fractional Quantum Anomalous {{Hall}} Effect in Multilayer Graphene},
  author = {Lu, Zhengguang and Han, Tonghang and Yao, Yuxuan and Reddy, Aidan P. and Yang, Jixiang and Seo, Junseok and Watanabe, Kenji and Taniguchi, Takashi and Fu, Liang and Ju, Long},
  year = {2024},
  journal = {Nature},
  volume = {626},
  number = {8000},
  pages = {759--764},
  publisher = {Nature Publishing Group},
  doi = {10.1038/s41586-023-07010-7}
}

@article{regan_mott_2020,
  title = {Mott and Generalized {{Wigner}} Crystal States in {{WSe2}}/{{WS2}} Moir{\'e} Superlattices},
  author = {Regan, Emma C. and Wang, Danqing and Jin, Chenhao and Bakti Utama, M. Iqbal and Gao, Beini and Wei, Xin and Zhao, Sihan and Zhao, Wenyu and Zhang, Zuocheng and Yumigeta, Kentaro and Blei, Mark and Carlstr{\"o}m, Johan D. and Watanabe, Kenji and Taniguchi, Takashi and Tongay, Sefaattin and Crommie, Michael and Zettl, Alex and Wang, Feng},
  year = {2020},
  journal = {Nature},
  volume = {579},
  number = {7799},
  pages = {359--363},
  publisher = {Nature Publishing Group},
  doi = {10.1038/s41586-020-2092-4}
}

@article{xie_tunable_2025,
  title = {Tunable {{Fractional Chern Insulators}} in {{Rhombohedral Graphene Superlattices}}},
  author = {Xie, Jian and Huo, Zihao and Lu, Xin and Feng, Zuo and Zhang, Zaizhe and Wang, Wenxuan and Yang, Qiu and Watanabe, Kenji and Taniguchi, Takashi and Liu, Kaihui and Song, Zhida and Xie, X. C. and Liu, Jianpeng and Lu, Xiaobo},
  year = {2025},
  journal = {arXiv:2405.16944},
  doi = {10.48550/arXiv.2405.16944},
  volume = {},
  pages = {}
}

@article{nam_lattice_2017,
  title = {Lattice Relaxation and Energy Band Modulation in Twisted Bilayer Graphene},
  author = {Nam, Nguyen N. T. and Koshino, Mikito},
  year = {2017},
  journal = {Phys. Rev. B},
  volume = {96},
  number = {7},
  pages = {075311},
  doi = {10.1103/PhysRevB.96.075311}
}

@article{tarnopolsky_origin_2019,
  title = {Origin of {{Magic Angles}} in {{Twisted Bilayer Graphene}}},
  author = {Tarnopolsky, Grigory and Kruchkov, Alex Jura and Vishwanath, Ashvin},
  year = {2019},
  journal = {Phys. Rev. Lett.},
  volume = {122},
  number = {10},
  pages = {106405},
  doi = {10.1103/PhysRevLett.122.106405}
}

@article{wang_fractional_2024,
  title = {Fractional {{Chern Insulator}} in {{Twisted Bilayer MoTe2}}},
  author = {Wang, Chong and Zhang, Xiao-Wei and Liu, Xiaoyu and He, Yuchi and Xu, Xiaodong and Ran, Ying and Cao, Ting and Xiao, Di},
  year = {2024},
  journal = {Phys. Rev. Lett.},
  volume = {132},
  number = {3},
  pages = {036501},
  doi = {10.1103/PhysRevLett.132.036501}
}

@article{mao_transfer_2024a,
  title = {Transfer Learning Relaxation, Electronic Structure and Continuum Model for Twisted Bilayer {{MoTe2}}},
  author = {Mao, Ning and Xu, Cheng and Li, Jiangxu and Bao, Ting and Liu, Peitao and Xu, Yong and Felser, Claudia and Fu, Liang and Zhang, Yang},
  year = {2024},
  journal = {Commun. Phys.},
  volume = {7},
  number = {1},
  pages = {262},
  publisher = {Nature Publishing Group},
  doi = {10.1038/s42005-024-01754-y}
}

@articel{zhangUniversalMoireModelBuildingMethod2024a,
  title = {Universal {{Moir\'e-Model-Building Method}} without {{Fitting}}: {{Application}} to {{Twisted MoTe}}\$\_2\$ and {{WSe}}\$\_2\$},
  author = {Zhang, Yan and Pi, Hanqi and Liu, Jiaxuan and Miao, Wangqian and Qi, Ziyue and Regnault, Nicolas and Weng, Hongming and Dai, Xi and Bernevig, B. Andrei and Wu, Quansheng and Yu, Jiabin},
  year = {2024},
  number = {2411.08108},
  journal = {arXiv},
  doi = {10.48550/arXiv.2411.08108}
}

@article{jia_moire_2024a,
  title = {Moir{\'e} Fractional {{Chern}} Insulators. {{I}}. {{First-principles}} Calculations and Continuum Models of Twisted Bilayer {{MoTe2}}},
  author = {Jia, Yujin and Yu, Jiabin and Liu, Jiaxuan and {Herzog-Arbeitman}, Jonah and Qi, Ziyue and Pi, Hanqi and Regnault, Nicolas and Weng, Hongming and Bernevig, B. Andrei and Wu, Quansheng},
  year = {2024},
  journal = {Phys. Rev. B},
  volume = {109},
  number = {20},
  pages = {205121},
  doi = {10.1103/physrevb.109.205121}
}

@misc{yu_relaxation_2025,
  title = {Relaxation and {{Its Effects}} on {{Electronic Structure}} in {{Twisted Systems}}: {{An Analytical Perspective}}},
  author = {Yu, Junxi and Wang, Bingbing and Liu, Cheng-Cheng},
  year = {2025},
  number = {2509.13114},
  journal = {arXiv},
  doi = {10.48550/arXiv.2509.13114}
}

@article{cantele_structural_2020,
  title = {Structural Relaxation and Low-Energy Properties of Twisted Bilayer Graphene},
  author = {Cantele, Giovanni and Alf{\`e}, Dario and Conte, Felice and Cataudella, Vittorio and Ninno, Domenico and Lucignano, Procolo},
  year = {2020},
  journal = {Phys. Rev. Research},
  volume = {2},
  number = {4},
  pages = {043127},
  doi = {10.1103/PhysRevResearch.2.043127}
}

@article{lisi_observation_2021,
  title = {Observation of Flat Bands in Twisted Bilayer Graphene},
  author = {Lisi, Simone and Lu, Xiaobo and Benschop, Tjerk and {de Jong}, Tobias A. and Stepanov, Petr and Duran, Jose R. and Margot, Florian and Cucchi, Ir{\`e}ne and Cappelli, Edoardo and Hunter, Andrew and Tamai, Anna and Kandyba, Viktor and Giampietri, Alessio and Barinov, Alexei and Jobst, Johannes and Stalman, Vincent and Leeuwenhoek, Maarten and Watanabe, Kenji and Taniguchi, Takashi and Rademaker, Louk and {van der Molen}, Sense Jan and Allan, Milan P. and Efetov, Dmitri K. and Baumberger, Felix},
  year = {2021},
  journal = {Nat. Phys.},
  volume = {17},
  number = {2},
  pages = {189--193},
  publisher = {Nature Publishing Group},
  doi = {10.1038/s41567-020-01041-x}
}

@article{lucignano_crucial_2019,
  title = {Crucial Role of Atomic Corrugation on the Flat Bands and Energy Gaps of Twisted Bilayer Graphene at the Magic Angle},
  author = {Lucignano, Procolo and Alf{\`e}, Dario and Cataudella, Vittorio and Ninno, Domenico and Cantele, Giovanni},
  year = {2019},
  journal = {Phys. Rev. B},
  volume = {99},
  number = {19},
  pages = {195419},
  doi = {10.1103/PhysRevB.99.195419}
}

@article{kazmierczak_strain_2021,
  title = {Strain Fields in Twisted Bilayer Graphene},
  author = {Kazmierczak, Nathanael P. and Van Winkle, Madeline and Ophus, Colin and Bustillo, Karen C. and Carr, Stephen and Brown, Hamish G. and Ciston, Jim and Taniguchi, Takashi and Watanabe, Kenji and Bediako, D. Kwabena},
  year = {2021},
  journal = {Nat. Mater.},
  volume = {20},
  number = {7},
  pages = {956--963},
  doi = {10.1038/s41563-021-00973-w}
}

@article{liu_twisted_2024a,
  title = {Twisted {{Magnetic Van}} Der {{Waals Bilayers}}: {{An Ideal Platform}} for {{Altermagnetism}}},
  author = {Liu, Yichen and Yu, Junxi and Liu, Cheng-Cheng},
  year = {2024},
  journal = {Phys. Rev. Lett.},
  volume = {133},
  number = {20},
  pages = {206702},
  doi = {10.1103/PhysRevLett.133.206702}
}

@article{yu_general_2025,
  title = {General Electronic Structure Calculation Method for Twisted Systems},
  author = {Yu, Junxi and Qian, Shifeng and Liu, Cheng-Cheng},
  year = {2025},
  journal = {Phys. Rev. B},
  volume = {111},
  number = {7},
  pages = {075434},
  doi = {10.1103/PhysRevB.111.075434}
}

@article{lopesdossantos_graphene_2007a,
  title = {Graphene {{Bilayer}} with a {{Twist}}: {{Electronic Structure}}},
  author = {{Lopes~dos~Santos}, J. M. B. and Peres, N. M. R. and Castro~Neto, A. H.},
  year = {2007},
  journal = {Phys. Rev. Lett.},
  volume = {99},
  number = {25},
  pages = {256802},
  publisher = {American Physical Society},
  doi = {10.1103/PhysRevLett.99.256802}
}

@article{bistritzer_moire_2011,
  title = {Moir{\'e} Bands in Twisted Double-Layer Graphene},
  author = {Bistritzer, Rafi and MacDonald, Allan H.},
  year = {2011},
  journal = {Proc. Natl. Acad. Sci. U.S.A.},
  volume = {108},
  number = {30},
  pages = {12233-12237},
  doi = {10.1073/pnas.1108174108}
}

@article{wu_topological_2019,
  title = {Topological {{Insulators}} in {{Twisted Transition Metal Dichalcogenide Homobilayers}}},
  author = {Wu, Fengcheng and Lovorn, Timothy and Tutuc, Emanuel and Martin, Ivar and MacDonald, A. H.},
  year = {2019},
  journal = {Phys. Rev. Lett.},
  volume = {122},
  number = {8},
  pages = {086402},
  doi = {10.1103/PhysRevLett.122.086402}
}

@article{koshino_maximally_2018,
  title = {Maximally {{Localized Wannier Orbitals}} and the {{Extended Hubbard Model}} for {{Twisted Bilayer Graphene}}},
  author = {Koshino, Mikito and Yuan, Noah F. Q. and Koretsune, Takashi and Ochi, Masayuki and Kuroki, Kazuhiko and Fu, Liang},
  year = {2018},
  journal = {Phys. Rev. X},
  volume = {8},
  number = {3},
  pages = {031087},
  doi = {10.1103/PhysRevX.8.031087}
}

@article{marzariMaximallyLocalizedGeneralized1997,
  title = {Maximally Localized Generalized Wannier Functions for Composite Energy Bands},
  author = {Marzari, Nicola and Vanderbilt, David},
  year = {1997},
  month = nov,
  journal = {Phys. Rev. B},
  volume = {56},
  number = {20},
  pages = {12847--12865},
  publisher = {American Physical Society},
  doi = {10.1103/PhysRevB.56.12847},
  urldate = {2022-01-22}
}

@article{mostofiWannier90ToolObtaining2008,
  title = {Wannier90: A Tool for Obtaining Maximally-Localised Wannier Functions},
  shorttitle = {Wannier90},
  author = {Mostofi, A. A. and Yates, J. R. and Lee, Y.-S. and Souza, I. and Vanderbilt, D. and Marzari, N.},
  year = {2008},
  month = may,
  journal = {Comput Phys Commun},
  volume = {178},
  number = {9},
  pages = {685--699},
  issn = {00104655},
  doi = {10.1016/j.cpc.2007.11.016},
  urldate = {2021-10-02},
}

@article{souzaMaximallyLocalizedWannier2001,
  title = {Maximally Localized Wannier Functions for Entangled Energy Bands},
  author = {Souza, Ivo and Marzari, Nicola and Vanderbilt, David},
  year = {2001},
  month = dec,
  journal = {Phys. Rev. B},
  volume = {65},
  number = {3},
  pages = {035109},
  publisher = {American Physical Society},
  doi = {10.1103/PhysRevB.65.035109},
  urldate = {2022-01-22}
}

@article{kresse_efficient_1996,
	title = {Efficient iterative schemes for \textit{ab initio} total-energy calculations using a plane-wave basis set},
	volume = {54},
	issn = {0163-1829, 1095-3795},
	url = {https://link.aps.org/doi/10.1103/PhysRevB.54.11169},
	doi = {10.1103/PhysRevB.54.11169},
	number = {16},
	urldate = {2024-06-26},
	journal = {Phys. Rev. B},
	author = {Kresse, G. and Furthmüller, J.},
	year = {1996},
	pages = {11169--11186},
}

@article{jung_initio_2014,
  title = {{\emph{Ab Initio}} Theory of Moir\'e Superlattice Bands in Layered Two-Dimensional Materials},
  author = {Jung, Jeil and Raoux, Arnaud and Qiao, Zhenhua and MacDonald, A. H.},
  year = 2014,
  journal = {Phys. Rev. B},
  volume = {89},
  number = {20},
  pages = {205414},
  doi = {10.1103/PhysRevB.89.205414}
}

@article{zhang_polarizationdriven_2024,
  title = {Polarization-Driven Band Topology Evolution in Twisted {{MoTe2}} and {{WSe2}}},
  author = {Zhang, Xiao-Wei and Wang, Chong and Liu, Xiaoyu and Fan, Yueyao and Cao, Ting and Xiao, Di},
  year = {2024},
  journal = {Nat. Commun.},
  volume = {15},
  number = {1},
  pages = {4223},
  doi = {10.1038/s41467-024-48511-x}
}

@article{liu_dpmoire_2025,
  title = {{{DPmoire}}: A Tool for Constructing Accurate Machine Learning Force Fields in Moir\'e Systems},
  author = {Liu, Jiaxuan and Fang, Zhong and Weng, Hongming and Wu, Quansheng},
  year = {2025},
  journal = {npj Comput Mater},
  volume = {11},
  number = {1},
  pages = {248},
  doi = {10.1038/s41524-025-01740-0}
}

@article{liDeeplearningDensityFunctional2022,
  title = {Deep-Learning Density Functional Theory {{Hamiltonian}} for Efficient Ab Initio Electronic-Structure Calculation},
  author = {Li, He and Wang, Zun and Zou, Nianlong and Ye, Meng and Xu, Runzhang and Gong, Xiaoxun and Duan, Wenhui and Xu, Yong},
  year = {2022},
  journal = {Nat Comput Sci},
  volume = {2},
  number = {6},
  pages = {367--377},
  doi = {10.1038/s43588-022-00265-6}
}

@article{bao_deeplearning_2024,
  title = {Deep-{{Learning Database}} of {{Density Functional Theory Hamiltonians}} for {{Twisted Materials}}},
  author = {Bao, Ting and Xu, Runzhang and Li, He and Gong, Xiaoxun and Tang, Zechen and Fu, Jingheng and Duan, Wenhui and Xu, Yong},
  url = {http://arxiv.org/abs/2404.06449},
  year = {2024},
  journal = {arXiv}
}

@article{naik_ultraflatbands_2018,
  title = {Ultraflatbands and {{Shear Solitons}} in {{Moir}}\textbackslash 'e {{Patterns}} of {{Twisted Bilayer Transition Metal Dichalcogenides}}},
  author = {Naik, Mit H. and Jain, Manish},
  year = {2018},
  journal = {Phys. Rev. Lett.},
  volume = {121},
  number = {26},
  pages = {266401},
  publisher = {American Physical Society},
  doi = {10.1103/PhysRevLett.121.266401}
}

@article{devakulMagicTwistedTransition2021,
  title = {Magic in Twisted Transition Metal Dichalcogenide Bilayers},
  author = {Devakul, Trithep and Cr{\'e}pel, Valentin and Zhang, Yang and Fu, Liang},
  year = {2021},
  journal = {Nat Commun},
  volume = {12},
  number = {1},
  pages = {6730},
  doi = {10.1038/s41467-021-27042-9}
}

@article{suarezmorell_flat_2010,
  title = {Flat Bands in Slightly Twisted Bilayer Graphene: {{Tight-binding}} Calculations},
  author = {Su{\'a}rez Morell, E. and Correa, J. D. and Vargas, P. and Pacheco, M. and Barticevic, Z.},
  year = {2010},
  journal = {Phys. Rev. B},
  volume = {82},
  number = {12},
  pages = {121407},
  doi = {10.1103/PhysRevB.82.121407}
}

@article{tramblydelaissardiere_localization_2010,
  title = {Localization of {{Dirac Electrons}} in {{Rotated Graphene Bilayers}}},
  author = {Trambly De Laissardi{\`e}re, G. and Mayou, D. and Magaud, L.},
  year = {2010},
  journal = {Nano Lett.},
  volume = {10},
  number = {3},
  pages = {804--808},
  doi = {10.1021/nl902948m}
}

@article{wang_vaspkit_2021,
  title = {{{VASPKIT}}: {{A}} User-Friendly Interface Facilitating High-Throughput Computing and Analysis Using {{VASP}} Code},
  author = {Wang, Vei and Xu, Nan and Liu, Jin-Cheng and Tang, Gang and Geng, Wen-Tong},
  year = {2021},
  journal = {Computer Physics Communications},
  volume = {267},
  pages = {108033},
  doi = {10.1016/j.cpc.2021.108033}
}

@article{cao_initio_2021,
  title = {{\emph{Ab Initio}} Four-Band {{Wannier}} Tight-Binding Model for Generic Twisted Graphene Systems},
  author = {Cao, Jin and Wang, Maoyuan and Qian, Shi-Feng and Liu, Cheng-Cheng and Yao, Yugui},
  year = {2021},
  journal = {Phys. Rev. B},
  volume = {104},
  number = {8},
  pages = {L081403},
  doi = {10.1103/PhysRevB.104.L081403}
}

@article{xie_lattice_2023,
  title = {Lattice Distortions, Moir\'e Phonons, and Relaxed Electronic Band Structures in Magic-Angle Twisted Bilayer Graphene},
  author = {Xie, Bo and Liu, Jianpeng},
  year = {2023},
  journal = {Phys. Rev. B},
  volume = {108},
  number = {9},
  pages = {094115},
  doi = {10.1103/PhysRevB.108.094115}
}

@article{li_tbplas_2023,
  title = {{{TBPLaS}}: {{A}} Tight-Binding Package for Large-Scale Simulation},
  author = {Li, Yunhai and Zhan, Zhen and Kuang, Xueheng and Li, Yonggang and Yuan, Shengjun},
  year = 2023,
  journal = {Computer Physics Communications},
  volume = {285},
  pages = {108632},
  doi = {10.1016/j.cpc.2022.108632}
}

@software{Cole_Python_Tight_Binding_2025,
author = {Cole, Trey and Coh, Sinisa and Vanderbilt, David},
doi = {10.5281/zenodo.12721315},
license = {GPL-3.0-or-later},
month = nov,
title = {{Python Tight Binding (PythTB)}},
url = {https://zenodo.org/records/12721315},
version = {2.0.0},
year = {2025}
}

@article{ma_topological_2021,
	title = {Topological flat bands in twisted trilayer graphene},
	volume = {66},
	copyright = {https://www.elsevier.com/tdm/userlicense/1.0/},
	issn = {20959273},
	url = {https://linkinghub.elsevier.com/retrieve/pii/S2095927320306575},
	doi = {10.1016/j.scib.2020.10.004},
	number = {1},
	urldate = {2024-06-27},
	journal = {Science Bulletin},
	author = {Ma, Zhen and Li, Shuai and Zheng, Ya-Wen and Xiao, Meng-Meng and Jiang, Hua and Gao, Jin-Hua and Xie, X.C.},
	year = {2021},
	pages = {18--22},
}

@article{gao_tunable_2020,
	title = {Tunable {Layer} {Circular} {Photogalvanic} {Effect} in {Twisted} {Bilayers}},
	volume = {124},
	issn = {0031-9007, 1079-7114},
	url = {https://link.aps.org/doi/10.1103/PhysRevLett.124.077401},
	doi = {10.1103/PhysRevLett.124.077401},
	number = {7},
	urldate = {2024-06-26},
	journal = {Physical Review Letters},
	author = {Gao, Yang and Zhang, Yinhan and Xiao, Di},
	month = feb,
	year = {2020},
	pages = {077401},
}
\end{document}